\documentclass[aps,prc,preprint,groupedaddress]{revtex4-1}

\usepackage{color,graphicx}
\usepackage{array,amsmath,amssymb,pifont,mathrsfs,latexsym}
\usepackage[pdftex,colorlinks]{hyperref}

\newcommand{\nuc}[2]{\ensuremath{^{\text{#1}}\text{#2}}}

\begin{document}


\title{Influence of Transport Variables on Isospin Transport Ratios}


\author{D. D. S. Coupland}
\email[]{coupland@nscl.msu.edu}
\affiliation{National Superconducting Cyclotron Laboratory and Physics and Astronomy Department, Michigan State University, East Lansing, MI 48824, USA}
\affiliation{Joint Institute of Nuclear Astrophysics, Michigan State University, E. Lansing, MI 48824, USA}

\author{W. G. Lynch}
\email[]{lynch@nscl.msu.edu}
\affiliation{National Superconducting Cyclotron Laboratory and Physics and Astronomy Department, Michigan State University, East Lansing, MI 48824, USA}
\affiliation{Joint Institute of Nuclear Astrophysics, Michigan State University, E. Lansing, MI 48824, USA}

\author{M. B. Tsang}
\email[]{tsang@nscl.msu.edu}
\affiliation{National Superconducting Cyclotron Laboratory and Physics and Astronomy Department, Michigan State University, East Lansing, MI 48824, USA}
\affiliation{Joint Institute of Nuclear Astrophysics, Michigan State University, E. Lansing, MI 48824, USA}

\author{P. Danielewicz}
\email[]{danielewicz@nscl.msu.edu}
\affiliation{National Superconducting Cyclotron Laboratory and Physics and Astronomy Department, Michigan State University, East Lansing, MI 48824, USA}

\author{Yingxun Zhang}
\email[]{zhyx@ciae.ac.cn}
\affiliation{China Institute of Atomic Energy, P.O. Box 275 (10), Beijing 102413, P.R. China}


\date{\today}

\begin{abstract}
The symmetry energy in the nuclear equation of state affects many aspects of nuclear astrophysics, nuclear structure, and nuclear reactions.  Recent constraints from heavy ion collisions, including isospin diffusion observables, have started to put constraints on the symmetry energy below nuclear saturation density, but these constraints depend on the employed transport model and input physics other than the symmetry energy.  To understand these dependencies, we study the influence of the symmetry energy, isoscaler mean field compressibility and momentum dependence, in-medium nucleon-nucleon cross sections, and light cluster production on isospin diffusion within the pBUU transport code.  In addition to the symmetry energy, several uncertain issues strongly affect isospin diffusion, most notably the cross sections and cluster production.  In addition, there is a difference in the calculated isospin transport ratios, depending upon whether they are computed using the isospin asymmetry of either the residue or of all forward moving fragments.  Measurements that compare the isospin transport ratios of these two quantities would help place constraints on the input physics, such as the density dependence of the symmetry energy.
\end{abstract}


\maketitle


\section{Introduction}
The symmetry energy in the nuclear equation of state, which describes the energy cost associated with differing proton and neutron densities, is currently an active field of research. The density dependence of the symmetry energy plays a role in many different nuclear physics systems at vastly differing size scales.  In nuclear structure, it is intimately related to the neutron skin of neutron-rich nuclei \cite{typel01,chen10,steiner05,horowitz01}, and it influences the centroid of the Giant Dipole Resonance \cite{trippa08} as well as the relative strength of the Pygmy Dipole Resonance \cite{klimkiewicz07,carbone10}. In astrophysics, nearly every observable property of neutron stars is affected by the density dependence of the symmetry energy \cite{steiner05,lattimer01}, and through this, nucleosynthesis and explosion mechanisms of core-collapse supernovae.  It also influences the evolution of nuclear reactions, including the neutron-to-proton spectral ratio from central heavy ion collisions \cite{famiano06}, isospin diffusion/charge equilibration \cite{tsang04,liu07}, isospin fractionation \cite{muller95}, and the ratio of charged pion yields \cite{baoan01}.  Despite this, the density dependence, and even the value at saturation density of the symmetry energy is not well determined.  Significant constraints at and below saturation density have emerged only within the past few years \cite{tsang09}, and there is still much debate about any constraints extracted  above saturation density \cite{reisdorf07,ferini05,xiao09,feng10,leifels93,reisdorf97,russotto11}.

Heavy ion collisions (HIC) provide a unique opportunity to study the density dependence of the nuclear equation of state, since they are the only opportunity we have in the laboratory to create nuclear matter significantly above and below saturation density.  Indeed, this is how the equation of state (EoS) of symmetric nuclear matter was mapped out~\cite{dan02s}.  Many HIC observables sensitive to the symmetry energy have been identified and are reviewed in Ref. \cite{baoan08}.  The effects of the symmetry energy on these observables, however, must be separated from the effects of the symmetric EoS, in medium nucleon-nucleon cross sections, and cluster formation.

To extract quantitative information about the symmetry energy from nuclear reactions, experimental data needs to be compared with the predictions of dynamical transport models.  Transport models track the time evolution of a nuclear reaction, and may separately treat the nuclear EoS, both symmetric and asymmetric parts, nucleon-nucleon (NN) collision cross sections, and inelastic nucleon collisions that lead to cluster or pion production~\cite{bert88}.  Most of the current nuclear reaction transport models either rely on the Uehling-Uhlenbeck version of the Boltzmann equation (BUU), or on quantum molecular dynamics (QMD)~\cite{bert88,baoan08,tsang09,colonna10}.  Both models are semiclassical.  The BUU approach uses many pointlike test particles per nucleon to approximate the continuous phase-space density matrix, while the QMD approach follows individual nucleons expressed as wave packets with a finite, usually fixed, width.  These different treatments affect the calculated dynamics.  Nucleon-nucleon collisions rearrange whole nucleons in the QMD approach while only rearranging individual test particles in the BUU approach, leading to larger fluctuations and more restrictive Pauli blocking in QMD.  Additionally, fragments form in the QMD treatment due to N-body correlations caused by the overlapping wave packets, but in BUU instead form mostly from mean-field instabilities, and are thus suppressed for many test particles per nucleon~\cite{colonna10}.

Even within either of the two models, however, particular numerical implementations can be different.  Different parameterizations or assumptions can be made about most of the transport properties.  The purpose of this paper is to assess within one such code the effects of these different assumptions on isospin diffusion, an observable of the symmetry energy that has received a lot of attention.  Comparisons will also be made to previous results in the literature, where available.

\subsection{Isospin Diffusion}

Isospin diffusion occurs in nuclear collisions where the projectile and target nuclei differ in their relative isospin asymmetry.  Over the course of the reaction, the symmetry energy pushes the system toward isospin equilibrium.  Since the reaction takes place over a limited time span, determined by the beam energy, full isospin equilibrium is not reached. The degree of isospin equilibration is an observable that is sensitive to the strength of the symmetry energy.  Usually isospin diffusion is discussed for semi-peripheral collisions, where large portions of the initial nuclei are spectators.  As these projectile-like and target-like residues separate in the expansion phase of the reaction, a low density neck connects them, and it is through this neck that isospin diffusion occurs.  
Thus isospin diffusion in peripheral and semi-peripheral collisions is sensitive to the behavior of the symmetry energy at sub-saturation density.  To maximize the sensitivity to the symmetry energy, the beam energy is chosen to be in the tens of MeV per nucleon.

Let us briefly discuss simple qualitative expectations regarding the effects of different ingredients on isospin diffusion.  All other effects being equal, a stronger symmetry energy in the low-density region of the neck will lead to more isospin diffusion, while nuclear collisions will tend to slow down the diffusion by locally trapping the asymmetry.  In fact, Shi and Danielewicz derived an expression for the diffusion constant from an analytic solution to the BUU equations, and found it to be proportional to the strength of the symmetry energy and inversely proportional to the neutron-proton cross section~\cite{shi03}.  Collisions are not expected to be the driving force in the energy regime studied here because of Pauli blocking. Diffusion is expected to increase with decreasing impact parameter due to the larger overlap region and longer reaction time.  For central impact parameters, however, a heavy residue generally does not survive, and some diffusion observables may not be well defined at the beam energy we studied. 

In the case of two colliding nuclei, however, other aspects of the system dynamics may complicate the above simple picture.  The symmetric matter mean field is not expected to affect isospin diffusion directly, but changes that affect the stability of nuclei or favor the creation of intermediate mass fragments (IMFs) may affect the dynamics of the system enough to enhance or suppress diffusion processes.  Baran et al. point out that density gradients will also lead to isospin transport, causing the low-density neck to become preferentially neutron-rich~\cite{baran05}.  Some of this matter is not transferred between nuclei but is expelled as free nucleons or light fragments, decreasing the total asymmetry of the remaining residues.  Other effects may change the role of nucleon-nucleon collisions. Some collisions may be needed to kick nucleons into the momentum space corresponding to stable and unstable orbits about the other nucleus.  In-medium clustering effects may increase the nucleon-nucleon cross section by increasing the available phase space.    

Isospin diffusion was measured for the systems of \nuc{124}{Sn} +\nuc{112}{Sn} and \nuc{112}{Sn}+\nuc{124}{Sn} at beam energies of 50 MeV/nucleon at NSCL/MSU~\cite{tsang04, liu07}.  The symmetric reaction systems, \nuc{124}{Sn}+\nuc{124}{Sn} and \nuc{112}{Sn}+\nuc{112}{Sn} were also studied to remove the contribution from non-isospin diffusion effects.  This was done by utilizing the isospin transport ratio (also called the imbalance ratio):
\begin{equation}
  R_{i}=2\frac{X-(X_{124+124}+X_{112+112})/2}{X_{124+124}-X_{112+112}},
\label{eq:Ri}
\end{equation}
first introduced in Ref. \cite{rami00}.  The ratio~(\ref{eq:Ri}) normalizes the amount of diffusion observed through an isospin sensitive observable $X$ to the value in the symmetric systems, such that in the absence of diffusion $R_i$ = 1 for a neutron-rich projectile, and $R_i$ = -1 for a neutron-poor projectile.  Complete equilibration occurs near $R_i = 0$. The ratio also reduces the sensitivity to pre-equilibrium emission, although it does not remove it entirely~\cite{baran05}.  In the MSU data, the isoscaling parameter $\alpha$ and the ratio of light mirror nuclei were both used to construct isospin transport ratios.   Another advantage of this ratio is that observables that depend linearly on each other produce the same ratio, allowing for easier comparison between data and transport models.  Transport models are generally not able to realistically handle sequential decays of the residues. 
 However, these models can calculate the isospin asymmetry $\delta$ of the excited residues.  Statistical and dynamical calculations as well as experimental tests have shown that the isoscaling parameter $\alpha$ relates linearly to the $\delta$ of the excited residues, so the experimental and theoretical results may be directly compared through the isospin transport ratio~\cite{liu07}. It should be noted, however, that it is uncertain if the yields used to construct $\alpha$ always came from the decay of the residue, as opposed to the breakup of the neck.

The MSU data from Ref.~\cite{tsang04} has been analyzed by several groups using different models to extract constraints on low density behavior of the symmetry energy.  Two of the most prominent results come from Ref.~\cite{baoan05}, which used the BUU code IBUU04, and Ref.~\cite{tsang09}, which used the QMD code ImQMD05.  Even through the constraints established by these two groups overlap, as shown in Fig. 3 of Ref.~\cite{tsang09}, there are real differences in the simulation results.  Aside from differences between the Boltzmann and molecular dynamics models, physics quantities are parameterized differently in the codes.  The form of the in-medium NN cross section reduction in particular was quite different~\cite{zhang11}.  IBUU04 includes isovector and isoscaler momentum dependencies of the mean field, while ImQMD05 includes only isoscaler.  More importantly, ImQMD05 can produce complex nuclei from the neck region, while IBUU04 cannot.  The relative importance of these transport properties on isospin diffusion appears to vary between the two models, but direct comparisons are difficult because of the differing nature of the models and the implementation of the codes.

The goal of this paper is to explore the effect of transport quantities and assumptions on isospin diffusion within a BUU transport model, henceforth referred to as pBUU \cite{dan91,dan92,dan00}.  We explore many of the basic properties of the system, including the density dependence of the symmetry energy, the value of the symmetry energy at saturation density, the curvature of the symmetric part of the EoS, the momentum dependence of the symmetric EoS, and in-medium nucleon cross sections.  The pBUU model can optionally produce light clusters in the medium, which is a process not native to the BUU framework and may make the system behave more similarly to the ImQMD05 work.  Due to its complexity, the momentum dependence of the isovector mean field is not studied here.  A similar, though not as extensive, study has also been carried out using the ImQMD05 code~\cite{zhang11}.  

\section{Description of Model}

The pBUU code is of the class of Boltzmann-Uehling-Uhlenbeck models, which are one of the main types of dynamic transport models used to study nuclear reactions, particularly of heavy nuclei.  The model includes as degrees of freedom both stable and excited baryons, pions, and light nuclear clusters.  It calculates the time evolution of the Wigner transform of the semiclassical one-body density matrix, i.e. the semiclassical one-body phase space distribution of those particles, following a set of Boltzmann equations, modified to include Pauli blocking~\cite{bert88}.  Many test particles per nucleon are used to approximate the continuous phase space distribution.  The mean field is calculated self-consistently from the phase space distribution.  The symmetric matter EoS has been extended to include momentum dependence, which can account for non-local interactions such as the Fock term.  This code has been extensively tested against experimental data, including single particle distributions~\cite{dan91,dan92}, elliptic flow~\cite{dan00}, and stopping observables~\cite{dan02app}.  It has been used to constrain the high density behavior of the symmetric matter EoS~\cite{dan02s}, and was used in early isospin diffusion calculations to demonstrate the sensitivity of the isospin transport ratio to the density dependence of the symmetry energy~\cite{tsang04}.  

\subsection{Mean field}

The symmetry energy used in this code takes the form:

\begin{equation}
  E_{sym}(\frac{\rho}{\rho_{0}})=
  S_{kin}(\frac{\rho}{\rho_{0}})^{2/3}+
  S_{int}(\frac{\rho}{\rho_{0}})^{\gamma_{i}}
\label{eq:esym}
\end{equation}

\noindent
where the kinetic term arises from the Fermi motion of the nucleons, resulting in $S_{kin}=12.5 MeV$.  We will examine the variation of isospin diffusion with potential interaction term, varying both the value at saturation density ($S_{int}$) and the density dependence ($\gamma_{i}$).  The form of the symmetric matter EoS is described in Ref.~\cite{dan00} for both momentum independent (MI) and momentum dependent (MD) forms.  As was done there, we choose parameter values that provide a curvature K=210 MeV, near the currently accepted value of $231\pm5$ MeV~\cite{baoan08}, and K=380 MeV for comparison.  We will show that the stiffness (curvature) of the symmetric matter EoS does not affect isospin diffusion.  We also study the momentum dependent case that best fits the elliptic flow data in Ref.~\cite{dan00}, which is characterized by an effective mass $m^{*} = 0.7 m_{N}$.  Our study shows that the momentum dependence changes the dynamics of the reaction such that intermediate mass fragments are produced from the breakup of the neck and the interaction time is lengthened, both of which affect the isospin diffusion signal.

\subsection{In-medium Cross sections}
\label{sec:disc_xs}

The residual interactions between nucleons are represented by nucleon-nucleon collisions, which are parameterized separately from the mean field.  The cross sections of these collisions are known to be reduced in the nuclear medium compared to their free-space values \cite{westfall93}, but the form of this reduction is not yet established.  Starting from the free space cross section parameterization of Ref.~\cite{cug95}, we examine two forms of the in-medium cross section reduction.  The screened cross section is derived from the geometric reasoning that the geometric cross section radius should not exceed the inter-particle distance, and is implemented in the form:

\begin{equation}
\sigma = \sigma_{0}\tanh(\sigma_{free}/\sigma_{0})
\end{equation}
\begin{equation}
\sigma_{0} = y\rho^{-2/3}, y=0.85
\end{equation}

\noindent
This screened reduction is strongly dependent on the density, and is very much reduced compared to free-space cross sections.  Since this form has a maximum cross section $\sigma_{0}$, large cross sections have a larger fractional reduction than small cross sections.  The neutron-proton (np) cross section is thus reduced more than the neutron-neutron (nn) or proton-proton (pp) cross sections.

In contrast, the Rostock cross section is parameterized from the results of Brueckner-Hartree-Fock calculations \cite{schulze97,schnell98}, and is implemented in the form:

\begin{equation}
\sigma=\sigma_{free}\exp(-0.6\frac{\rho}{\rho_{0}}\frac{1}
{1+(T_{c.m.}/150MeV)^{2}})
\end{equation}

\noindent
where $T_{c.m.}$ is the center of mass kinetic energy of the nucleon pair.  It is less dependent on density than the screened cross section, and results in less reduction at the energy of interest.  Both cross sections were tested versus stopping data within the pBUU model in Ref. \cite{dan02app}. The screened cross section showed somewhat better agreement with data, but both were an improvement compared to free-space cross sections. Neither cross section reduction corresponds exactly with the reductions used in the IBUU04 or ImQMD05 codes, but the amount of reduction in those codes is more similar to the Rostock cross section than to the screened. We will show that the choice of cross section reduction can greatly affect the diffusion signal.

\subsection {Cluster Formation}

Production of light clusters in the pBUU code is implemented as an option up through cluster mass 3, with the production rate derived through the inverse process of cluster breakup.  In pBUU these clusters are produced by the interaction of test particles in the nuclear medium.  Since the clusters are test particles themselves, they may exist as part of the density distribution that represents a nucleus, residue, or fragment.  In this analysis, this makes them distinct from intermediate mass fragments (IMFs), which are defined as density distributions that are spatially separated from the residues and each other, and are the result of the interplay between density fluctuations and the mean field.  While this is far from a complete clustering model, missing in particular the formation of $\alpha$ particles, it can give an indication of the importance of clustering in the nuclear medium. Even though the QMD model includes a method of forming clusters, it is difficult to remove the cluster formation mechanism entirely from the QMD model in order to quantify the effects of clusters on the reaction dynamics.  With the cluster option in pBUU, we are able to compare isospin diffusion results with and without cluster formation.


\section{Identification of Residue}

Several methods of extracting the projectile-like residue from the results of a transport simulation have been put forward in the literature.  These methods generally fall into two categories.  Within the first method, employed exclusively in BUU approaches providing single-particle density, the residue is defined as consisting of cells in the computational grid that match a particular average velocity cut and average density cut, regardless of the cells' relative locations~\cite{baoan05,tsang04}.  Within the second method, spatially contiguous fragments are identified, either by examining the single-particle density directly or through a tree-spanning method.  Thereafter, cuts on the average velocity and size of those fragments are applied to either include only the largest fragment (the residue), or all fragments that match particular mass and velocity criteria, sometimes including free nucleons~\cite{tsang09,zhang11,rizzo08}.  In both cases, the velocity cut tends to be half of the initial projectile velocity in the center of mass, and if a density cut is employed, it can range from 0.2 to 0.05 of saturation density. Obviously, these two methods will produce the same result for a simple reaction that does not produce any intermediate mass fragments.  Nonetheless, the first method has the advantage that it allows fragments to be traced through the timescale of the reaction, while the second can only define fragments once the system has sufficiently spread out.  On the other hand, the second method allows for a more precise description of the final state of the system when intermediate mass fragments form, and it does it not break up a spatially contiguous fragment if some rotation or other collective motion exists within that fragment.  Whatever method is chosen to define the projectile-like residue, the isospin asymmetry can be determined and used as the isospin sensitive observable within the isospin transport ratio.

In our analysis, we chose to use the second method, since there are systems produced in this study that emerge either with IMFs or collective motion in the residue regions. We identified contiguous areas above a given density threshold as fragments, and we tested the effect of the density cut, of the time chosen to stop the reactions, and of velocity cuts on the extracted isospin asymmetry and isospin transport ratio, using several combinations of physics inputs examined here.  The density cut was varied from $\rho_{0}/20$ to $\rho_{0}/5$.  Lower density cuts led to a more asymmetric residue, indicating a low-density neutron skin on the fragment, but this effect was almost entirely removed by the isospin transport ratio. Likewise, after the main reaction is completed (when all fragments have spatial separation), the residues continue to evaporate neutrons, becoming more symmetric, but this effect exists in the symmetric systems as well and ultimately does not affect the isospin transport ratio significantly.  The time for the fragments to separate spatially depended on the density cut and on the physics inputs, but was complete by 220 fm/c after the start of the simulation in every case.  For the remainder of this paper, the analysis was conducted at 270 fm/c with a density cut of $\rho_{0}/20$, unless otherwise indicated.

Some combinations of physics inputs yielded results that were sensitive to velocity cuts.  As will be discussed later, some choices of mean field parameters affect the formation of intermediate mass fragments (IMF).  Systems that produced large IMFs sometimes had significantly different results for the isospin transport ratio if those large fragments were included in the analysis.  These effects will be discussed in more detail in the next section.  Systems without IMFs or with many small IMFs showed no dependence on whether the highest mass fragment was considered alone or whether all fragments matching a particular velocity cut were considered together, for any forward-moving velocity cut from 0 to 1/2 of the initial velocity of the projectile in the center of mass.  For the remainder of this paper, our analysis will focus on either the asymmetry of the largest forward-moving fragment without IMFs, or on the asymmetry of all forward-moving fragments.  Free nucleons are not included in the analysis.

\section{Results}

\subsection{Mean field effects}

\begin{figure}
  \begin{center}
    \includegraphics[width=6.0cm]{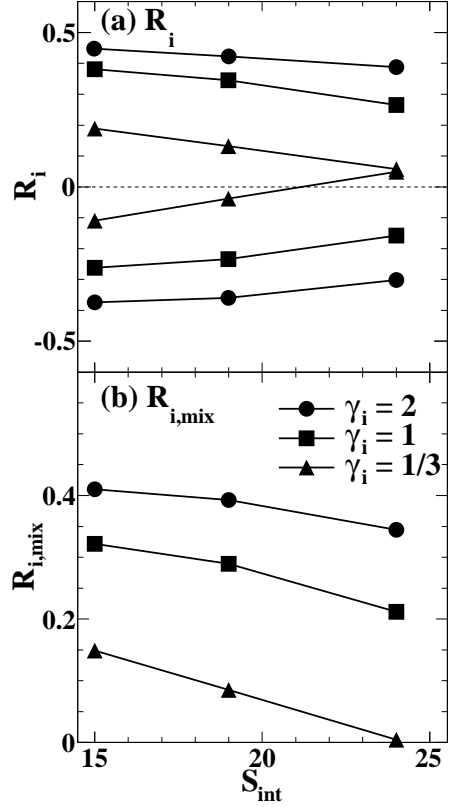}
  \end{center}
  \caption{(a) The isospin transport ratio of the \nuc{124}{Sn} +\nuc{112}{Sn} and \nuc{112}{Sn}+\nuc{124}{Sn} systems as a function of the symmetry energy interaction coefficient, for a momentum independent (MI) mean field. (b)  Results from combining the two systems.}
  \label{fig:Mi_Sint_K380}
\end{figure}

We start with a simple momentum independent (MI) description of the mean field.  As has been done in Ref.~\cite{dan00}, we use a mean field with an isoscaler compressibility of K=380 MeV, a screened cross section, and no light cluster production.  We vary the coefficient ($S_{int}$) and density dependence coefficient ($\gamma_{i}$) of the symmetry energy in Eq.~\ref{eq:esym}.  The results when following the heavy residue are shown in Fig.~\ref{fig:Mi_Sint_K380}(a)
for the \nuc{124}{Sn} +\nuc{112}{Sn} (positive $R_{i}$) and \nuc{112}{Sn}+\nuc{124}{Sn} (negative $R_{i}$) systems at an impact parameter of 6 fm.  The x-axis represents the strength of the interaction term, $S_{int}$, in Eq.~\ref{eq:esym} at saturation density. Three different density dependencies $\gamma_{i}$ = 1/3, 1, and 2 are shown with triangles, squares, and circles.  Lines are drawn to guide the eye.  For all $\gamma_{i}$, larger $S_{int}$ leads to an isospin transport ratio closer to zero (horizontal dotted line), which matches the expectation that a stronger symmetry energy leads to more diffusion and faster isospin equilibration.  Consistent with that expectation, the most equilibration occurs for the smallest value of $\gamma_{i}$ (triangles), which corresponds to a stronger symmetry energy at sub-saturation densities.  This confirms that isospin diffusion is sensitive to the strength of the symmetry energy in the sub-saturation neck region, as anticipated.

The isospin transport ratios for the \nuc{124}{Sn} +\nuc{112}{Sn} and \nuc{112}{Sn}+\nuc{124}{Sn} systems converge for $\gamma=1/3$ and $S_{int}=24$, indicating complete equilibration, but the value of $R_i$ at these points is definitely different from zero.  This offset can be attributed to the mass asymmetry of the system, as may be demonstrated with a very simple calculation.  Assuming that nucleons are only transfered between the nuclei and none are lost to free space or to light fragments, simple algebra shows that complete equilibration leads to $R_{i,eq}=(A_{1}-A_{2})/(A_{1}+A_{2})$.  This result will be modified by preequilibrium emission, but the effect on the  transport ratio will be small.  In our system this calculation leads to $R_{i,eq}=0.051$, which is consistent with the observed equilibration in Fig.~\ref{fig:Mi_Sint_K380}a.  However, $R_{i,eq}=0$ is more intuitive, and noting that the two systems provide similar information, we can restore the expectation of a zero for complete equilibration by combining information from both systems:

\begin{equation*}
  R_{i,mix} = (R_{i,124+112} - R_{i,112+124})/2
\end{equation*}

\noindent
as demonstrated in Fig.~\ref{fig:Mi_Sint_K380}(b).  Unless stated otherwise, we use the quantity $R_{i,mix}$ for the remainder of this paper, and for simplicity we relabel it as $R_i$.  Having established that varying either $S_{int}$ or $\gamma_{i}$ produces changes resulting only from the overall strength of the symmetry energy at subsaturation densities, for the calculations discussed in the remainder of the paper we vary only $\gamma_{i}$ and set $S_{int} = 19 MeV$ ($E_{sym}(\rho_{0}) = 31.5 MeV$), as for the middle set of points in Fig.~\ref{fig:Mi_Sint_K380}.

Next we 
\begin{figure}
  \begin{center}
    \includegraphics[width=6.0cm]{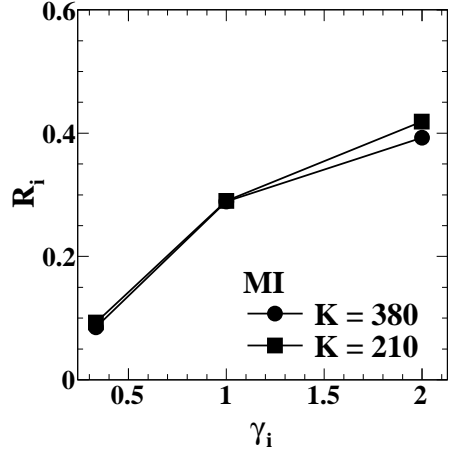}
  \end{center}
  \caption{The effect of the compressibility of the symmetric EoS on isospin diffusion, as a function of the density dependence of the symmetry energy.}
  \label{fig:Mi_K_mix}
\end{figure}
examine the effect of changing the isoscaler compressibility, still using a momentum independent mean field.  In Fig.~\ref{fig:Mi_K_mix}, the simulations  with compressibility K=210 MeV are shown with squares and K=380 MeV with circles, as a function of the density dependence of the symmetry energy.  Recall that the accepted value of K is $231\pm5$ MeV~\cite{baoan08}, so K=380 MeV lies well outside the accepted uncertainties.  However, varying the compressibility even this much produces very little change in the isospin transport ratio, although there is slightly less diffusion for the K=210 MeV case than for the K=380 MeV case.  

\begin{figure}
  \begin{center}
    \includegraphics[width=6.0cm]{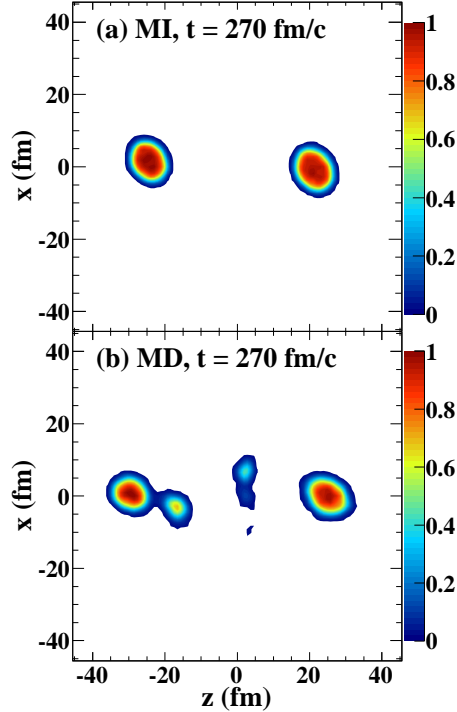}
  \end{center}
  \caption{(Color online) Density profiles from sample calculations within the reaction plane at t=270fm/c, after the residues have separated, following an EoS that is (a) momentum independent and (b) momentum dependent.  The z axis is the beam direction, and the color scale represents density, normalized to saturation density.}
  \label{fig:Mi_vs_Md_prof}
\end{figure}

In elliptic flow data at higher energies, including momentum dependence had an effect similar to increasing the stiffness of the mean field.  That is why the momentum dependent (MD) mean field with curvature K=210 MeV was often compared to the momentum independent (MI) mean field with  K=380 MeV~\cite{dan00}.  However, for isospin diffusion at 50 MeV/nucleon, the momentum dependence has a much different effect than the compressibility.  First, it  changes the dynamics of the reaction, enhancing the likelihood of IMF production, and in particular leading to the production of larger IMFs.  This arises because the depth of the attractive potential is lower at large relative momenta, and it is easier for groups of particles to escape.  A density profile of the momentum independent and the momentum dependent systems is shown in Fig.~\ref{fig:Mi_vs_Md_prof} at t=270 fm/c, highlighting the difference in IMF production.  

\begin{figure}
  \begin{center}
    \includegraphics[width=6.0cm]{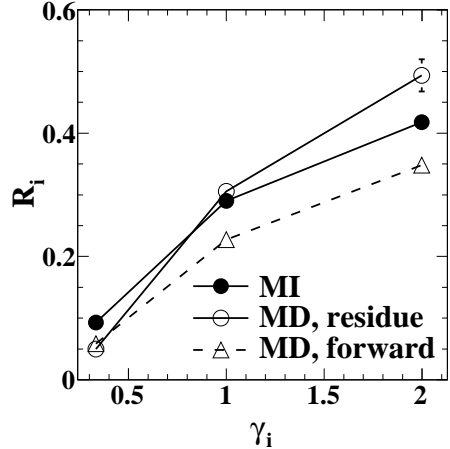}
  \end{center}
  \caption{The effect of momentum dependence on isospin diffusion, as a function of the density dependence of the symmetry energy.  The diffusion into the largest residue is compared to the diffusion into all forward-moving fragments.}
  \label{fig:Mi_vs_Md}
\end{figure} 

The emerging IMFs have a large effect on the diffusion signal.  
When all forward-moving fragments are used as the isospin tracer, including both the projectile-like residue and IMFs, a MD EoS gives rise to more diffusion than a MI EoS (Fig.~\ref{fig:Mi_vs_Md}, open triangles and filled circles, respectively), resulting in smaller values of $R_{i}$.  Only very small IMFs are produced in MI simulations.  The isospin diffusion results are essentially unchanged when the small IMFs are included, so the respective result is not shown separately in the figure.
By contrast, when looking just at the largest residue, inclusion of momentum dependence (open circles) decreases the amount of diffusion, resulting in larger $R_i$ values for large $\gamma_{i}$ and smaller for small $\gamma_{i}$.  
Thus, inclusion of the momentum dependence of the mean field has two effects on the diffusion signal.  It increases the overall rate of isospin diffusion, 
but if the overall rate of diffusion is slow, as is the case for large $\gamma_{i}$, many of the diffused particles are stuck in the neck and contribute to the IMFs rather than to the heavy residue.  If the diffusion rate is fast, as is the case for small $\gamma_{i}$, the  neck and the residue reach equilibrium, and the inclusion of IMFs does not change the diffusion signal.  This is similar to an effect of IMFs reported by Zhang \emph{et al.} in Ref. \cite{zhang11} following the ImQMD code, although the dependence on $\gamma_{i}$ is different there.  

\begin{figure}
  \begin{center}
    \includegraphics[width=6.0cm]{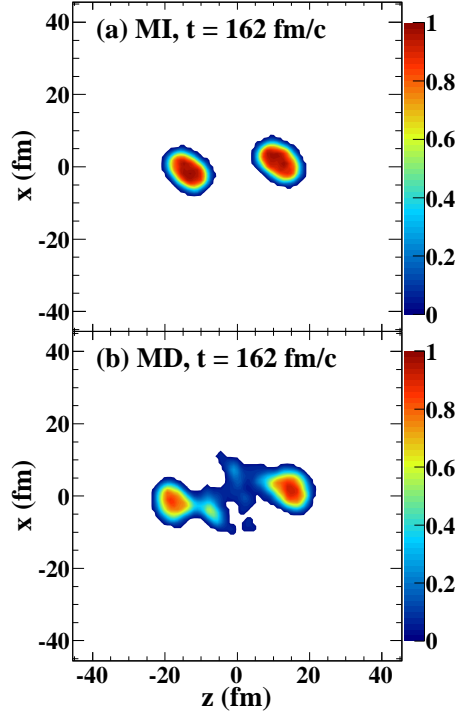}
  \end{center}
  \caption{(Color online) Density profiles within the reaction plane at t=162fm/c, normalized to saturation density.  (a) When following a momentum independent (MI) EoS and  (b) when following a momentum dependent (MD) EoS. }
  \label{fig:Mi_vs_Md_prof_162}
\end{figure}

On the other hand, the increase in diffusion due to the momentum dependence contradicts the results of Rizzo \emph{et al.} in Ref. \cite{rizzo08}, who report the opposite effect within the stochastic mean field (SMF) model.  They report that the inclusion of momentum dependence in the mean field, which causes greater isoscaler repulsion, increases the rate at which the residues separate and thus reduces the interaction time.  The momentum dependence corresponds to a reduction of the effective mass compared to the free mass, which means that particles with a given canonical momentum will be moving faster in a momentum dependent mean field than in a momentum independent mean field. This speeds up the interaction of the colliding nuclei.  Our simulations do show that the residues move apart from each other more quickly when isoscaler momentum dependence is included.  However, the formation of density clumps in the neck, which results in emission of IMFs after the neck snaps, anchor the neck, causing it to persist longer and allowing more diffusion to occur.  This effect is visible in the density profiles of the system at 162 fm/c, when the neck in the MI case (upper panel) has already broken up but the MD case (lower panel) has not, as shown in Fig. \ref{fig:Mi_vs_Md_prof_162}.
This effect occurs even when the density fluctuations in the neck do not persist as IMFs at the end of the reaction.  This suggests that the major effect of the symmetric mean field in isospin transport is to influence the time and manner of the neck breakup. The explicit inclusion of stochastic effects in the SMF model produces density fluctuations leading to IMFs even for MI mean fields, which may cause the neck to break up similarly regardless of the momentum dependence of the mean field.  The amount of time during which the residues interact would then depend only on the velocity of the residues, producing the different trend in their calculations.

\begin{figure}
  \begin{center}
    \includegraphics[width=6.0cm]{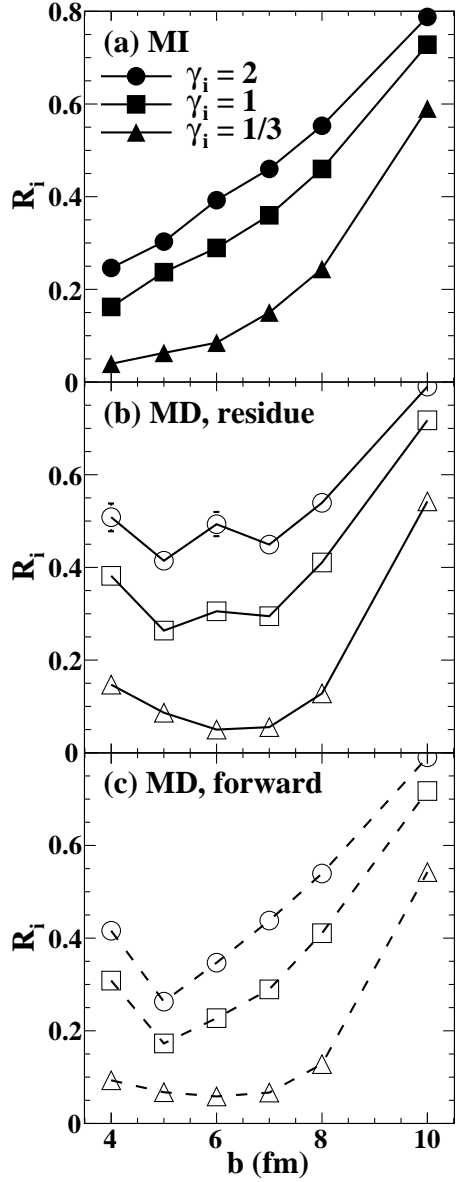}
  \end{center}
  \caption{Impact parameter dependence of isospin diffusion for: (a) momentum independent EoS,  (b) momentum dependent EoS, to the heavy residue,  (c) momentum dependent EoS, to all forward-moving fragments.}
  \label{fig:impactparam}
\end{figure}

This same dependence on manner of the neck breakup is evident in the impact parameter dependence of the isospin transport ratio.  Fig.~\ref{fig:impactparam}
shows the transport ratio as a function of impact parameter. Line style corresponds to density dependence of the symmetry energy.  As expected, less diffusion takes place (larger $R_{i}$) at large impact parameters where the overlap region is smaller and the residues continue to move faster.  The amount of diffusion changes monotonically in the MI case (Fig.~\ref{fig:impactparam}(a)).  When  momentum dependence is included, however, this monotonicity disappears (Fig.~\ref{fig:impactparam}(b)) for central and mid-peripheral collisions.  This is again due to the large IMFs, which are only created if there is a sufficient overlap region.  For peripheral collisions ($b > 7 fm$), the density fluctuations in the neck extend the lifetime of the neck, increasing the interaction time and the amount of equilibration, but not creating large enough IMFs to carry away the asymmetry.  For mid-peripheral collisions ($5 fm \leq b \leq 7 fm$), the residue equilibration varies significantly but flattens out on average. When including all forward moving fragments (Fig.~\ref{fig:impactparam}(c)), the monotonicity is restored in this region.  For the smallest  impact parameter ($b = 4 fm$) shown in Fig.~\ref{fig:impactparam}, the amount of equilibration indicated by both tracers decreases (larger $R_i$ values).  In this region, the motion of matter squeezed out from the overlap region competes with the diffusion process.  Comparison of the three panels reveals the smallest effect from the momentum dependence of the mean field at large impact parameters, while still being sensitive to $\gamma_{i}$.  For mid-peripheral collisions, the dependence on the mechanism of neck breakup due to the momentum dependence of the mean field is reduced when looking at all forward-moving fragments.

\subsection{In-medium cross sections}
\label{sec:results:xs}

We now move in our considerations from the mean field to the in-medium nucleon cross sections.  As discussed in section \ref{sec:disc_xs}, we examine free cross sections and two parameterizations of in-medium reductions.  Prior to this point in the paper, we have been using the screened cross section.

\begin{figure}
  \begin{center}
    \includegraphics[width=6.0cm]{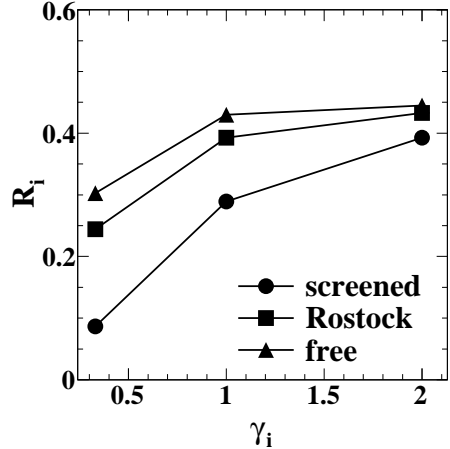}
  \end{center}
  \caption{Effect of different cross section parameterization on isospin diffusion.}
  \label{fig:Mi_xs}
\end{figure}

The effect of the different cross section parameterizations is shown in Fig.~\ref{fig:Mi_xs},
 when following the MI EoS.  Clearly, less diffusion occurs for the free cross section (triangles) than for either reduced one, and this effect is most pronounced for small $\gamma_i$.  This contradicts the conclusion arrive at within the IBUU04 code in Ref. \cite{baoan05}, where it was found that the free cross section produced more diffusion than the reduced cross section for very stiff symmetry energies ($\gamma_i > 2$), with less effect in the range of $\gamma_i$ considered here.  Similarly, the ImQMD05 code is largely insensitive to the cross section over a similar range of density dependencies as considered here~\cite{zhang11}.  However, both of those results were obtained with MD mean fields, which will be considered here only later in Fig. \ref{fig:Md_xs}.  In Fig. \ref{fig:Mi_xs}, the difference between the Rostock and screened cross sections, shown as squares and circles, respectively, is larger than the difference between the free and the Rostock.  More importantly, the sensitivity to $\gamma_i$ is strongly affected by the cross section.  

\begin{figure}
  \begin{center}
    \includegraphics[width=6.0cm]{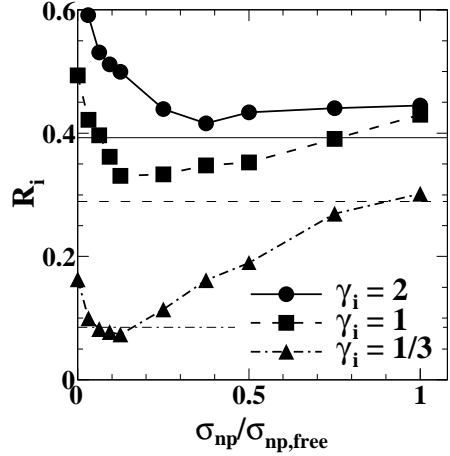}
  \end{center}
  \caption{Sensitivity of isospin diffusion to np cross section reduction by a constant factor.  The x-axis is the fractional reduction relative to free cross sections.  Horizontal lines represent the results obtained using the screened cross section, from Fig.~\ref{fig:Mi_Sint_K380}.}
  \label{fig:npmult}
\end{figure}

\begin{figure}
  \begin{center}
    \includegraphics[width=6.0cm]{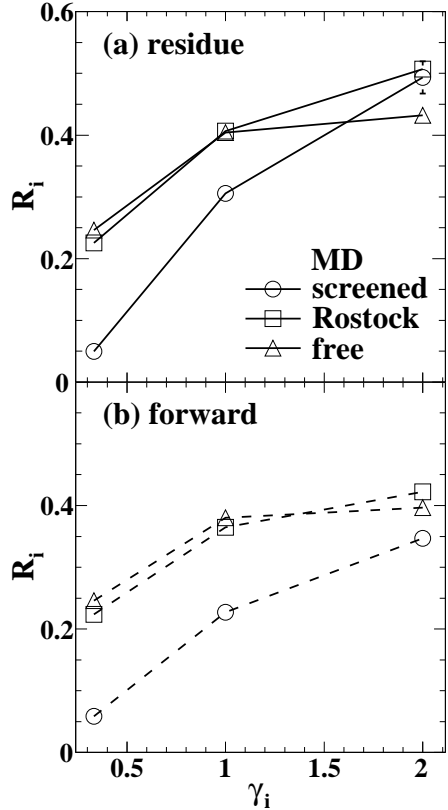}
  \end{center}
  \caption{Isospin equilibration for a momentum dependent EoS with in-medium cross section reductions. Panel (a) shows results from the heavy residues only, while panel (b) shows results from all forward-moving fragments}
  \label{fig:Md_xs}
\end{figure}

The screened and Rostock cross sections lead to similar stopping in heavy ion collisions because they produce similar viscosities~\cite{dan02app}.  However, they suppress the overall number of nucleon-nucleon collisions by different factors, and as discussed in section~\ref{sec:disc_xs}, the screened cross section reduces the np cross section more than the nn or pp.  In these simulations, $N_{Rostock} \approx \frac{3}{4}N_{free}$ while $N_{nn,pp,screened} \approx \frac{1}{4}N_{free}$ and $N_{np,screened} \approx 0.15 N_{free}$, where $N$ is the total number of nucleon-nucleon collisions over the course of the simulation.  It is not simply the number of collisions that affects the isospin transport.  From Shi and Danielewicz~\cite{shi03}, we expect the reduction in the np cross section to strongly influence the isospin diffusion rate.  Fig.~\ref{fig:npmult}
 shows the isospin transport values when the np cross section is reduced by a constant factor relative to the free cross sections.  As expected, changing the nn and pp cross sections leads to only very small changes, so these are not shown.  The x-axis is the employed reduction factor, with the limit of no collisions at the left of the axis ($\sigma_{np}/\sigma_{np,free} = 0$) and free-space cross sections at the right.  Line and marker styles represent the density dependence of the symmetry energy, $\gamma_i$.  The trend of each line indicates that collisions have two major effects on the amount of isospin diffusion.  For nearly free cross sections on the right side of the  plot, increasing the cross section decreases the amount of diffusion.  This matches the intuition, formally derived in Ref. \cite{shi03}, that collisions impede isospin diffusion.  A nucleon that undergoes a collision or series of collisions with large momentum transfer forgets its original direction of motion, which impedes the equilibration effects of the mean field.  

For the very small cross sections on the left side of the figure, however, increasing the cross section increases diffusion.   The nucleon-nucleon collisions can cause isospin transport  by  knocking nucleons out of the momentum space of one nucleus, causing them to be transferred to the other nucleus or expelled as free nucleons.  This process is largely isospin independent, but will contribute to isospin equilibration via simple mixing.  The hook at low end of the $\gamma_{i}=1/3$ (triangles, dot-dashed) line shows that this yields a finite contribution to isospin diffusion for even a strong symmetry potential at subsaturation densities.  For $\gamma_i=2$ (solid line), this is the dominant contribution, and is approaching the point where diffusion would increase by increasing the cross section at any point on the curve.  For even stiffer symmetry energies, this could easily lead to the free cross section producing more diffusion, as was seen in the IBUU04 study.  For reference, three horizontal lines are drawn to indicate the value of $R_i$ from the screened cross section.  Except for the $\gamma_{i}=1/3$ case (dot-dashed), a uniform reduction of the free cross section never produces as much diffusion as the screened cross section.  This indicates that the strong density dependence of the screened cross section is particularly favorable for overall isospin equilibration.  Therefore, the exact form of the in-medium cross section is the important quantity, not merely the  viscosity or the net collision number.

A similar hierarchy of cross section effects is evident with a momentum dependent EoS, with the stipulation that one takes the IMFs into account.  As shown in Fig.~\ref{fig:Md_xs},
 the IMFs included in the forward moving fragments modify the signal differently for the different cross sections.  The screened cross section (circles) is strongly affected by the inclusion of IMFs (panel (b)), while the Rostock (squares) is affected modestly and the free cross section (triangles) is affected little.  This may open an avenue to experimentally constrain the cross sections.  A careful comparison of Fig.~\ref{fig:Md_xs} to Fig.~\ref{fig:Mi_xs} also reveals that the results obtained using the screened cross section are most strongly modified by using a mean field with momentum dependence.  This can be understood as the competition between the collisions and the mean field:  as the importance of the collisions grows, the changes in the mean field become less important. This figure also shows that with a momentum dependent mean field, the Rostock and free cross sections produce very similar results, except for the very stiff symmetry energy $\gamma_i = 2$.  This is consistent with the results of \cite{baoan05}.

\begin{figure}
  \begin{center}
    \includegraphics[width=6.0cm]{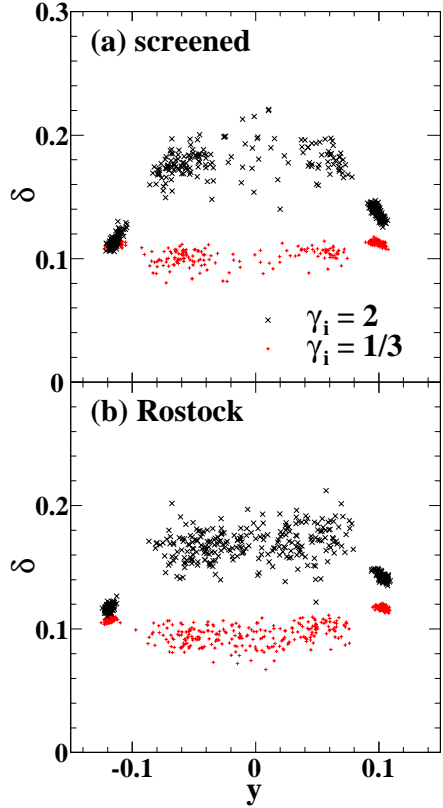}
  \end{center}
  \caption{(Color online) Distribution of fragments in isospin asymmetry and rapidity, using (a) screened and (b) Rostock in-medium cross sections.}
  \label{fig:IMFrap}
\end{figure}

The difference in dynamics that gives rise to the different IMF dependencies is evident in Fig.~\ref{fig:IMFrap}, where the asymmetry of all fragments greater than mass two at the final time of the \nuc{124}{Sn} +\nuc{112}{Sn} simulations are plotted against their rapidity in the reaction center of mass, accumulated over many simulations.
The two different symbols represent two different $\gamma_i$ values in the EoS used in the simulations, and panels (a) and (b) show  results for the screened and Rostock cross sections, respectively.  The tight clumps of fragments at the largest absolute rapidities are the large residues, and all other points represent IMFs.  While the fragment asymmetry is similar between the two cross sections for a given $\gamma_i$, the IMFs are more evenly spread in rapidity with the Rostock cross section (Fig.~\ref{fig:IMFrap}(b)), whereas they are more likely to be concentrated at larger absolute rapidity with the screened cross section (Fig.~\ref{fig:IMFrap}(a)).  This indicates less stopping for the IMFs from the screened cross section, even though the residue rapidity is unchanged.  Related to this, but not illustrated, is that the produced IMFs tend to be larger in the screened case compared to the Rostock case, but of quite similar asymmetry.  
Note that there are no IMFs close to the residue in rapidity.  Since this is a snapshot taken on the timescale of the nuclear reaction, the effect cannot be attributed to the long-term impact of Coulomb repulsion.  Rather it is the effect of the residue ``gobbling up'' nearby fragments.  Because the IMFs are moving faster in the screened case, they are closer to the residue and are more likely to be absorbed.  This, combined with the mass difference of the IMFs, causes the different degree of response to the inclusion of IMFs in the isospin transport ratio.  The dependence on IMF mass and rapidity distributions can be measured, and this may provide additional constraints on the in-medium cross section reduction.

\subsection{Cluster Production}

\begin{figure}
  \begin{center}
    \includegraphics[width=6.0cm]{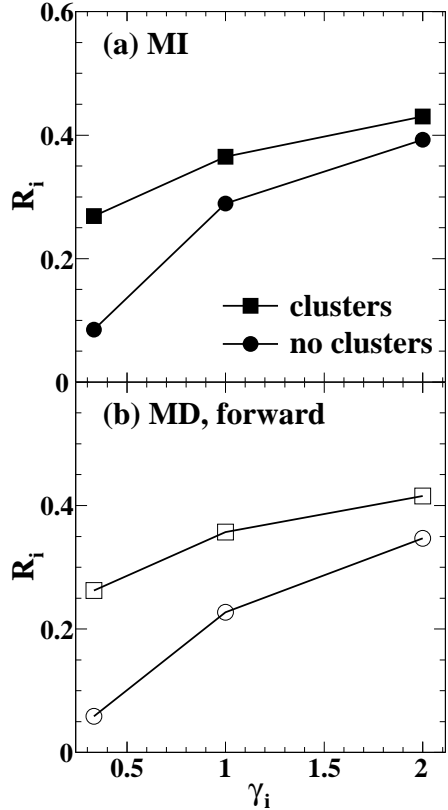}
  \end{center}
  \caption{Effect of in-medium cluster production on isospin equilibration for a momentum independent mean field (a) and a momentum dependent mean field (b). }
  \label{fig:clusters}
\end{figure}

\begin{figure}
  \begin{center}
    \includegraphics[width=6.0cm]{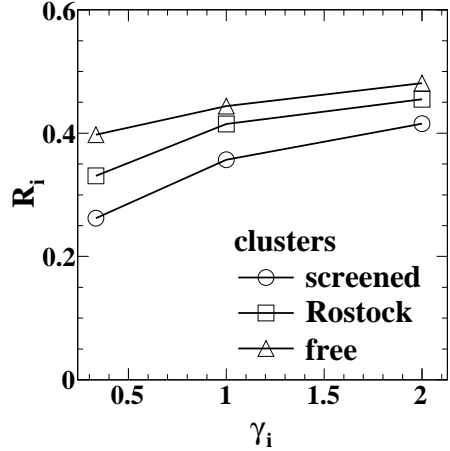}
  \end{center}
  \caption{Effect of cluster production and of in-medium cross section reductions on isospin equilibration, when using a momentum dependent mean field.}
  \label{fig:clusters_xs}
\end{figure}

Next we examine the effects of light cluster production in the nuclear medium.  This is a distinct process from the formation of IMFs during the breakup of the neck.  IMFs are the result of density fluctuations, whereas cluster production is a fast nucleation process, the result of inelastic particle collisions.  It is unclear how to combine the information about light fragments produced from these two mechanisms; nevertheless, it is interesting to examine the change in dynamics that results. Fig.~\ref{fig:clusters}
 shows the changes in the isospin transport ratio caused by including cluster production, when examining all forward moving fragments.  Specifically, the figure compares the results of simulations with and without clustering when employing a MI mean field (panel (a)), and when employing a MD mean field (panel (b)).  In both cases, the inclusion of cluster production (squares) reduces the amount of diffusion and decreases the sensitivity of the diffusion to the symmetry energy.  The increased value of the isospin transport ratio for small $\gamma_i$ better matches the predictions of ImQMD05~\cite{zhang11}, bolstering the idea that clustering in a BUU model is required for comparison to results from QMD models and to experimental data.  However, the maximum isospin transport ratio for large $\gamma_i$ here is still significantly less than that reported in ImQMD05.  The effect of clustering is very similar to the effect of increasing the cross section in section~\ref{sec:results:xs}.  In fact, with this new collision channel open, more collisions occur.  This reduces the influence of other effects.  Comparing the two panels in Fig.~\ref{fig:clusters}, there is little difference between the momentum dependent and independent simulations, once clustering is included.  Similarly, the differences between cross section parameterizations are reduced when clustering is included, compare Figs.~\ref{fig:clusters_xs} and~\ref{fig:Md_xs}.  The difference between the residue and forward moving fragment tracers is also less with the inclusion of cluster production, and for the free and Rostock cross sections, it becomes a small, nearly constant offset in the value of the isospin transport ratio.  Interestingly, the simple dependence on collisions number, that was lost when momentum dependence was included, has been restored.  

\begin{figure}
  \begin{center}
    \includegraphics[width=6.0cm]{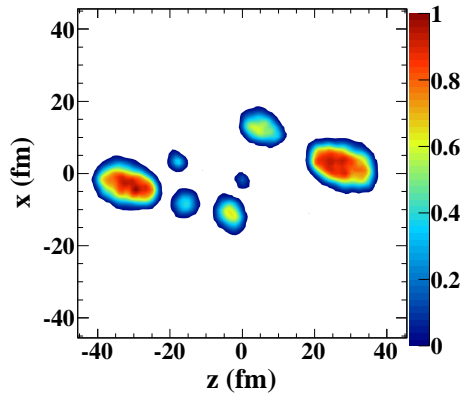}
  \end{center}
  \caption{(Color online) Density profile within the reaction plane at t=270fm/c, normalized  to saturation density, when cluster production is incorporated and a momentum-dependent mean field is employed.}
  \label{fig:prof_Md_clus}
\end{figure}

\begin{figure}
  \begin{center}
    \includegraphics[width=6.0cm]{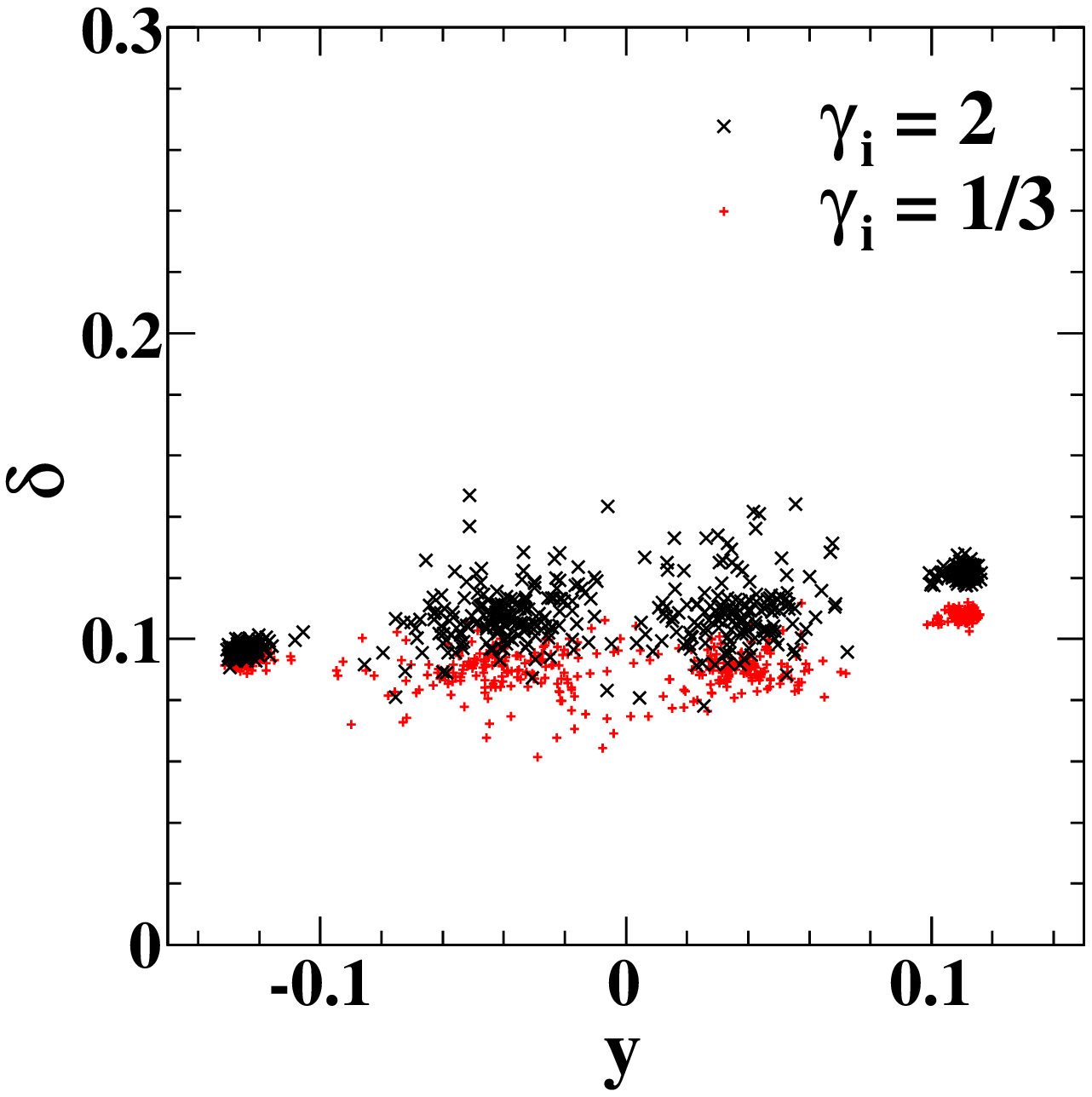}
  \end{center}
  \caption{(Color online) Distribution of fragments in isospin asymmetry and rapidity, when cluster production is incorporated and a momentum dependent mean field is employed, for two choices of symmetry energy.}
  \label{fig:fragrap_Md_clus}
\end{figure}

However, the effect of including clustering goes well beyond the increased number of collisions.
  The extra kinetic energy released by the cluster formation causes the neck to fragment into many smaller pieces rather than a few large IMFs, as seen in the density plot at t=270 fm/c in Fig.~\ref{fig:prof_Md_clus}.  The fragmentation process also continues longer, as evidenced by the shape of the residues, which have not yet reached a spherical shape.  Additionally, the whole neck region tends to expel its asymmetry with less dependence on $\gamma_i$, as shown in Fig.~\ref{fig:fragrap_Md_clus} (compare Fig. \ref{fig:IMFrap}).  This effect was predicted by Shi and Danielewicz \cite{shi00}. They argued that the clusters modify the liquid-gas phase transition, forming droplets in the low-density gas phase and causing it to behave more like the liquid phase.  This results in less isospin migration to the neck, and thus less diffusion taking place through the neck. 
Another result is that the IMFs are smaller and have asymmetries similar to the residue, and both these effects reduce the importance of including them in the isospin transport ratio.  Note that the isospin distribution of the IMFs in Fig.~\ref{fig:fragrap_Md_clus} is similar to that in Fig.~\ref{fig:IMFrap} for $\gamma_i=1/3$, while the diffusion represented in Fig.~\ref{fig:clusters_xs} is more similar to that previously displayed for $\gamma_i=2$.  This indicates that the effect of clustering on isospin diffusion may be distinguished from the effect of the symmetry energy by examining both the IMFs and the projectile-like residues.

\begin{figure}
  \begin{center}
    \includegraphics[width=6.0cm]{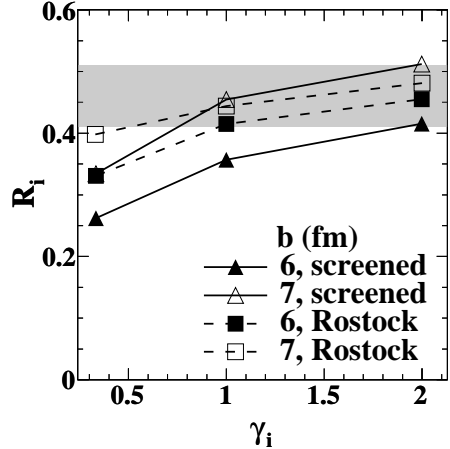}
  \end{center}
  \caption{A comparison of isospin equilibration data to results of simulations that incorporate a MD mean field, cluster production, and one of two parameterizations of in-medium cross section reductions.}
  \label{fig:datacompare}
\end{figure}

Finally, it is interesting to consider a comparison to data, with all the discussed effects included.  Fig.~\ref{fig:datacompare} includes simulation results using a momentum dependent mean field, cluster production, and both cross section reductions, compared to the experimental data (shaded region) from Ref.~\cite{tsang04}.  The simulations are carried out at impact parameters of both 6fm (solid symbols) and 7fm (open symbols), to represent the uncertainty in the impact parameter in the data.  The results compare well with the data, but unfortunately do not tightly constrain the density dependence of the symmetry energy, $\gamma_i$ especially when the Rostock cross sections (squares) are used.  This calls for further investigation into the uncertainties in the  transport model.  As presented here, many of those uncertainties affect the distribution of mass, isospin, and rapidity between IMFs and heavy residues, which can be experimentally measured.  The decreased dependence on $\gamma_i$ once all effects are included, as compared to previous BUU results, also argues for more precise experimental measurements.  Isospin diffusion experiments have been proposed at both the National Superconducting Cyclotron Laboratory and the Rare Isotope Beam Facility at RIKEN to measure the isospin signals from both IMFs and heavy residues with high precision.

\section{Conclusion}

We have used the pBUU transport model code to explore the effect of various aspects of the input physics and transport description on isospin diffusion.  Consistent with previous studies, we found diffusion to be more significant if the symmetry energy is larger at subsaturation densities.  This can occur when the symmetry energy is more weakly dependent on density.  Turning our attention to the influence of other transport quantities, we found some unexpected sensitivities.  For example, we found that the inclusion of momentum dependence in the isoscaler mean field produced density fluctuations in the neck, prolonging the diffusion process and giving rise to IMFs.  In addition, the choice of in-medium NN cross section can also affect isospin diffusion, producing both a change in the IMF distribution and altering the balance between a mean-field driven diffusion process and a collision-driven mixing.  Depending on the magnitude of the in-medium cross section and density dependence of the mean field, we found that nucleon-nucleon cross sections can enhance diffusion, for very small cross sections, or reduce diffusion, for large cross sections comparable to the values in free space.  This study examined for the first time the effect of including in-medium cluster production on the isospin diffusion process. Previous studies have looked at the effect of IMF formation, but have not looked at in-medium cluster production in a way similar to pBUU.  We found that cluster production reduced diffusion by causing the neck to become more isospin symmetric. 

In all cases that produce IMFs, considering either the residue alone or all forward-moving fragments yield different isospin diffusion signals, with the residue values for the isospin transport ratio consistent with less diffusion.  These can be considered two observables of isospin diffusion, each sensitive to the input physics, and they can both be exploited to improve constraints on the symmetry energy from comparisons between data and calculations.  To do this, however, it is important to have experimental data which originates unambiguously from the projectile residue, and this is not the case in the published isospin diffusion data.  New experiments aim to measure and compare the different diffusion signals from IMFs and large residues, and this should help to constrain the effects studied here. 

The large effects observed in this study suggest that more work is needed to pin down the input physics in transport models other than the symmetry energy. Similar studies will be needed on other observables, both within the same code and between different codes and different models, employing standard parameterizations of the input physics for appropriate comparison.  Observables of the symmetry energy are sensitive enough to the collision dynamics that the differences between BUU and QMD dynamics need to be understood and accounted for. To constrain all of these sensitive transport quantities and conclusively determine the value of the symmetry energy at subsaturation density, it would be helpful to explore and obtain a consistent map of many observables in a multiparameter physics input space.

\begin{acknowledgments}
This work has been supported by the U.S. National Science Foundation under grants PHY-0216783, 0606007, and 0800026. We wish to acknowledge the support of the Michigan State University High Performance Computing Center and the Institute for Cyber Enabled Research.
\end{acknowledgments}

%

\end{document}